\documentclass{aa}

\usepackage{graphics}
\begin{document}

\thesaurus{06(08.09.2 Gliese 866; 08.02.6; 08.12.2; 03.20.2)}

\title{The nearby M-dwarf system \object{Gliese~866} revisited
\thanks{Based on observations collected at the German-Spanish Astronomical
 Center on Calar Alto, Spain, and at the European Southern Observatory,
 La Silla, Chile}}
\author{J.\,Woitas\inst{1}
  \and Ch.\,Leinert\inst{1}
  \and H.\,Jahrei{\ss}\inst{2}
  \and T.\,Henry\inst{3}
  \and O.\,G.\,Franz\inst{4}
  \and L.\,H.\,Wasserman\inst{4}}
\offprints{J.\,Woitas, e-mail: woitas@mpia-hd.mpg.de}

\institute{Max-Planck-Institut f\"ur Astronomie, K\"onigstuhl 17,
  D-69117 Heidelberg, Germany
\and Astronomisches Rechen-Institut, M\"onchhofstr. 12-14, D-69120 Heidelberg,
  Germany
\and Harvard-Smithsonian Center for Astrophysics, Cambridge, MA 02138-1516,
 USA
\and Lowell Observatory, Flagstaff, AZ 86001-4499, USA}
\date{Received 24 August 1999 / accepted 12 October 1999}

\maketitle

\begin{abstract}
We present an improved orbit determination for the visual pair in the M-dwarf
triple system \object{Gliese~866} that is based on new speckle-interferometric
and HST observations. The system mass is $M = 0.34\pm 0.03\,M_{\odot}$.
The masses of the components derived using the mass-luminosity relation are
consistent with this mass sum. All three components of \object{Gliese 866}
seem to have masses not far from the hydrogen burning mass limit.

\end{abstract}

\section{Introduction}
It is now a well established fact that there exists a large number of
substellar objects (see e.\,g. the recent review article by Oppenheimer et al.
\cite{Oppenheimer00}). This increases the need for a better understanding of
stellar properties at the lower end of the main sequence. Nearby low-mass
stellar systems are particularly important for this purpose, because their
orbital motion allows dynamical mass determinations.\\
Here we will consider the nearby triple system \object{Gliese~866}
(Other designations: LHS 68, WDS 22385-1519). Using
speckle interferometry, Leinert et al.\,(\cite{Leinert86}) and also
McCarthy et al.\,(\cite{McCarthy87}) discovered a companion (henceforth
\object{Gliese~866}~B) located about 0\farcs4 away from the main
component \object{Gliese~866}~A. Following observations proved the
possibility to cover the whole orbit of this binary system within a
few years by speckle-interferometric observations. Based on 16 data
points, Leinert et al.\,(\cite{Leinert90}, hereafter L90) presented a first
determination of orbital parameters and masses. They derived the
combined mass of the system to be $0.38\pm 0.03\,M_{\odot}$. This was
inconsistent with values of $M_{\mathrm{A}} \approx 0.14\,M_{\odot}$ and
$M_{\mathrm{B}} \approx
0.11\,M_{\odot}$ obtained from empirical mass-luminosity relations and
stellar interior models (see L90 and references therein).\\
In the meantime an additional spectroscopic companion (hereafter called
\object{Gliese~866}~a) to \object{Gliese~866}~A has been detected
(Delfosse et al.\,\cite{Delfosse99}). This is a plausible reason for the
mentioned mass excess in \object{Gliese~866}, but -- given the fact that only
$0.38\,M_{\odot}$ now had to be distributed among three stars -- raises the
question for a substellar component in the \object{Gliese~866} system.\\
To improve on the mass determination of the components of 
\object{Gliese~866} we have taken 20 more speckle-interferometric
observations and one additional HST determination of relative position.
These show the orbit of the wide pair: B with respect to Aa. We present an
overview of the observations and the data reduction process in
Sect.\,\ref{obs}. The results are given in Sect.\,\ref{results},
discussed in Sect.\,\ref{discussion} and are summarized in
Sect.\,\ref{summary}.  

\section{Observations and data reduction}
\label{obs}
\begin{figure*}
\resizebox{12cm}{!}{\includegraphics{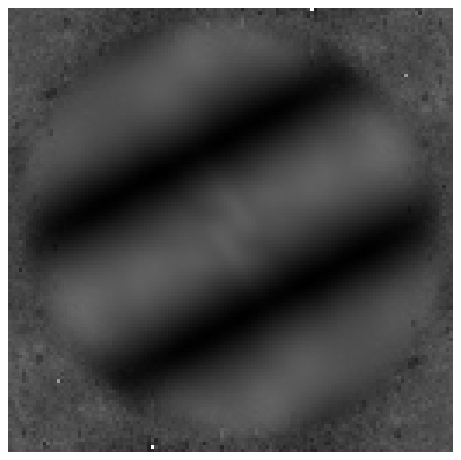}\includegraphics{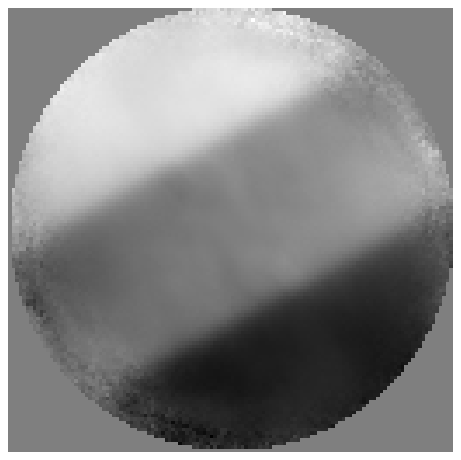}}
\hfill
\parbox[b]{55mm}{
\caption{\label{vispha} Modulus (left) and bispectrum-phase (right) of
 the complex visibility for \object{Gliese~866}. They have been computed from
 speckle-interferometric data obtained on 27 September 1996 at the
 3.5\,m-telescope on Calar Alto in the K-band using the NIR array camera
 MAGIC. The bispectrum-phase has only been calculated inside a circle that
 corresponds to the Nyquist spatial frequency which is
 $7.0\,\mathrm{arcsec}^{-1}$ for the adopted pixel scale.}}
\end{figure*}

A list of the new observations and their results is given in
Table\,\ref{speckle-results}. The observations numbered 1, 3, 5 and 7
used one-dimensional speckle-interferometry. Details of observational
techniques and data reduction for this method are described in
Leinert \& Haas\,(\cite{Leinert89}).\\
\begin{table*}
\caption{\label{speckle-results} Results of new observations of
 \object{Gliese~866}. Position angle and projected separation
 are those of \object{Gliese~866}~B relative to \object{Gliese~866}~Aa.
 The abbreviations in column ''Telescope'' denote Calar Alto (CA), Kitt Peak
 National Observatory (KPNO), European Southern Observatory (ESO) and
 New Technology Telescope (NTT). The latter has an aperture of 3.5\,m.}
\begin{tabular}{lllllllll}
& & & & & & & &\\ \hline
No. & Date & Epoch & Position & Projected
 & Filter & $I_{\mathrm{B}}/I_{\mathrm{Aa}}$ & Telescope & Camera \\
 & & & angle [$\degr$] & separation & & & (see caption) & \\
 & & &                 & [mas] & & & & \\ \hline
1 & 01.09.1990 & 1990.6681 & 348.0 $\pm$ 2.2 & 481 $\pm$ 18 & K &
 0.58 $\pm$ 0.03 & CA 3.5\,m & 1D \\
2 & 04.12.1990 & 1990.9254 & 339.8 $\pm$ 2.4 & 463 $\pm$ 19 & K & & KPNO 3.8\,m
 & 2D \\
3 & 19.09.1991 & 1991.7173 & 186.5 $\pm$ 3.4 & 186 $\pm$ 11 & K &
 0.55 $\pm$ 0.02 & CA 3.5\,m & 1D \\
4 & 29.10.1991 & 1991.8268 & 155.5 $\pm$ 3.8 & 200 $\pm$ 13 & K &
 0.51 $\pm$ 0.03 & CA 3.5\,m & 1 - 5\,$\mu$m\\
5 & 16.05.1992 & 1992.3751 & 15.3 $\pm$ 0.5 & 244 $\pm$ 6 & K &
 0.57 $\pm$ 0.05 & ESO 3.6\,m & 1D \\
6 & 14.10.1992 & 1992.7885 & 350.8 $\pm$ 1.5 & 449 $\pm$ 11 & K &
  0.58 $\pm$ 0.01 & CA 3.5\,m & 1 - 5\,$\mu$m \\
7 & 10.01.1993 & 1993.0273 & 348.6 $\pm$ 1.4 & 480 $\pm$ 12 & K &
 0.55 $\pm$ 0.03 & CA 3.5\,m & 1D \\
8 & 29.07.1993 & 1993.5750 & 319.3 $\pm$ 1.2 & 259 $\pm$ 11 & K &
 0.618 $\pm$ 0.071 & ESO NTT & SHARP \\
9 & 02.10.1993 & 1993.7529 & 277.7 $\pm$ 0.5 & 105 $\pm$ 4 & K &
 0.60 $\pm$ 0.03 & CA 3.5\,m & MAGIC \\
10 & 01.05.1994 & 1994.3313 & 74.9 $\pm$ 2.7 & 126 $\pm$ 4 & K &
 0.562 $\pm$ 0.031 & ESO NTT & SHARP \\
11 & 14.09.1994 & 1994.7036 & 5.9 $\pm$ 0.4 & 327 $\pm$ 4 & K  &
 0.552 $\pm$ 0.009 & CA 3.5\,m & MAGIC \\
12 & 12.12.1994 & 1994.9473 & 354.0 $\pm$ 0.2 & 434 $\pm$ 4 & K &
 0.551 $\pm$ 0.017 & CA 3.5\,m & MAGIC \\
13 & 13.12.1994 & 1994.9501 & 354.3 $\pm$ 0.3 & 439 $\pm$ 4 & 917\,nm &
  0.603 $\pm$ 0.009 & CA 3.5\,m & MAGIC \\
14 & 09.07.1995 & 1995.5201 & 337.3 $\pm$ 0.3 & 439 $\pm$ 4 & K &
 0.581 $\pm$ 0.014 & ESO NTT & SHARP \\
15 & 16.08.1995 & 1995.6243 & 333.1 $\pm$ 0.3 & 392 $\pm$ 4 & K &
 0.587 $\pm$ 0.022 & ESO 3.6\,m & SHARP\,2 \\
16 & 08.06.1996 & 1996.4381 & 131.6 $\pm$ 0.2 & 156 $\pm$ 1 & 583\,nm &
    0.69 $\pm$ 0.06 & HST & FGS3 \\
17 & 27.09.1996 & 1996.7419 & 28.0 $\pm$ 0.5 & 188 $\pm$ 4 & K &
 0.568 $\pm$ 0.022 & CA 3.5\,m & MAGIC \\
18 & 25.08.1997 & 1997.6489 & 340.2 $\pm$ 0.2 & 485 $\pm$ 4 & K &
 0.593 $\pm$ 0.006 & ESO 3.6\,m & SHARP\,2 \\
19 & 16.11.1997 & 1997.8761 & 333.3 $\pm$ 0.1 & 391 $\pm$ 4 & K &
 0.58 $\pm$ 0.01 & CA 3.5\,m & MAGIC \\
20 & 07.05.1998 & 1998.3477 & 214.8 $\pm$ 0.3 & 104 $\pm$ 7 & K &
 0.558 $\pm$ 0.004 & ESO NTT & SHARP \\
21 & 10.10.1998 & 1998.7748 & 101.2 $\pm$ 0.2 & 133 $\pm$ 4 & K &
 0.497 $\pm$ 0.026 & CA 3.5\,m & OMEGA Cass \\ \hline
\end{tabular}
\end{table*}
All other speckle observations were done using two-dimensional infrared array
cameras. Sequences of typically 1000 images with exposure
times of $\approx 0.1\,\mathrm{sec}$ were taken for \object{Gliese~866} and a
nearby reference star. After background subtraction, flatfielding and badpixel
correction these data cubes are Fourier-transformed.\\
We determine the modulus of the complex visibility (i.\,e. the Fourier
transform of the object brightness distribution) from power spectrum
analysis. The phase is recursively reconstructed using two different
methods: The Knox-Thompson algorithm (Knox \& Thompson\,\cite{Knox74}) and
the bispectrum analysis (Lohmann et al.\,\cite{Lohmann83}). Modulus and phase
are characteristic strip patterns for a binary. As an example we
show them in Fig.\,\ref{vispha} for the observation done at 27 September 1996
on Calar Alto. By fitting a binary model to the complex visibility we derive
the binary parameters: position angle, projected separation and flux ratio.\\
To obtain a highly precise relative astrometry which is crucial for orbit
determination one has to provide a good calibration of pixel scale and
detector orientation. For the speckle observations since 9 July 1995 this
calibration has been done using astrometric fits to images of the Trapezium
cluster, where precise astrometry has been given by McCaughrean \& Stauffer 
(\cite{McCaughrean94}). During the previous observing runs
binary stars with well known orbits were observed for calibrating pixel scale
and detector orientation. By doing subsequent observations of these systems
and calibrating them  with the Trapezium cluster we have put all speckle
observations since July 1993 in a consistent system of pixel scale and
detector orientation.

\section{Results}
\label{results}
\begin{figure}
\resizebox{\hsize}{!}{\includegraphics{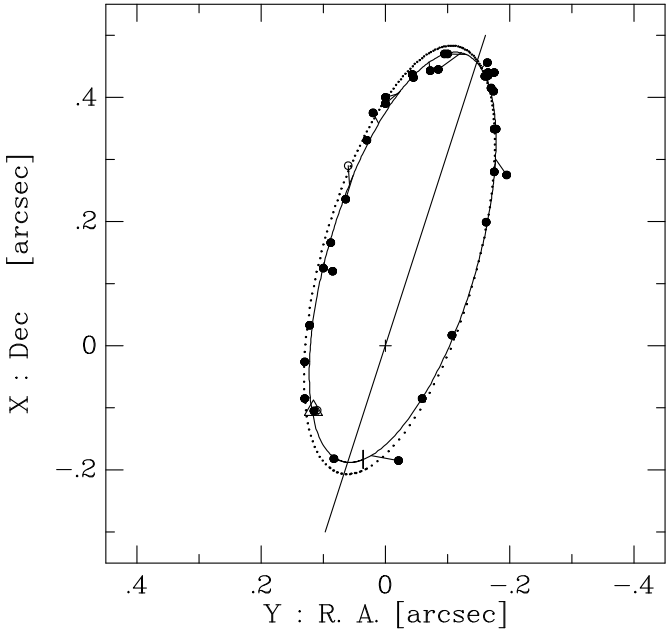}}
\caption{\label{gl866-orbit} Improved relative orbit of \object{Gliese~866}~AB
 (solid) compared to the orbit derived by L90 (dotted).
 North is up and east is to the left.
 The individual measurements are connected to the predicted positions
 by lines. For the two measurements
 represented by open circles only one coordinate could be derived from
 the observation, the other one is taken from the predicted orbit. The triangle
 denotes the HST measurement (No.\,16 in Table\,\ref{speckle-results}).
 The periastron is indicated by a vertical bar. The line of nodes also is
 plotted.}
\end{figure}
After combining the relative astrometry from L90 and our new observations,
there are now 37 independent data points for the orbital motion of the visual
pair in \object{Gliese~866}. They are plotted in Fig.\,\ref{gl866-orbit},
together with the result of an orbital fit that used the method of
Thiele and van den Bos, including iterative differential corrections
(Heintz \cite{Heintz78}).
\begin{table}
\caption{\label{bahnparams} Orbital elements for \object{Gliese\,866}}
\begin{tabular}{ll}
& \\ \hline
Orbital element & Result \\ \hline
Node &  $\Omega = 162.1^{\circ}\pm 0.4^{\circ}$ \\
Longitude of periastron &  $\omega = -17.7^{\circ}\pm 1.1^{\circ}$ \\ 
Inclination & $i = 112.4^{\circ}\pm 0.5^{\circ}$ \\
Semi majoraxis & $A = (0.346\pm 0.004)\,\mathrm{arcsec}$ \\
Period & $P = (2.2506\pm 0.0033)\,\mathrm{yr}$ \\
Eccentricity & $e = 0.437\pm 0.007$ \\
Periastron time &  $t_P = 1987.236\pm 0.014$ \\ \hline
\end{tabular}
\end{table}
The orbital elements resulting from a fit to the full data set are given in
Table\,\ref{bahnparams}. Since we don't know which node is ascending and which
one is descending, we choose -- as usually is done -- $\Omega$ to be between
$0\degr$ and $180\degr$. $\omega$ is the angle between the adopted
$\Omega$ and the periastron (positive in the direction of motion), and
$i > 90\degr$ means clockwise motion.\\
\object{Gliese~866} has a trigonometric parallax
$\pi = 289.5\pm 4.4\,\mathrm{mas}$ (van Altena et al.\,\cite{vanAltena95}).
In the following calculations we use the external error
of the parallax: $6.8\,\mathrm{mas}$ instead of the (internal) value given by
van Altena et al. This yields a distance of $3.45\pm 0.08\,\mathrm{pc}$,
a semi majoraxis of $1.19\,\mathrm{AU}$ and finally -- using
Kepler's third law -- a system mass $M_{\mathrm{Sys}} = 0.336\pm
0.026\,M_{\odot}$.
This result remains unchanged within the uncertainties if the calculation is
done with natural subsets of the data (see Table\,\ref{subsets}). There is
particularly no significant difference if only the 2D data points
with good astrometric calibration (see Sect.\,\ref{obs}) are used.
Because it covers the longest time span, we take the result for the full
data set as best values for orbit and system mass.

\begin{table}
\caption{\label{subsets} Derived system mass for different subsets of
 the data}
\begin{tabular}{lll}
& & \\ \hline
Data set & N & $M_{\mathrm{Sys}}/M_{\odot}$ \\ \hline
all data & 37 & 0.336 $\pm$ 0.026 \\
only our speckle data & 28 & 0.336 $\pm$ 0.032 \\
only our 2D speckle data with & 13 & 0.340 $\pm$ 0.035 \\
direct astrometric calibration & & \\
\hline
\end{tabular}
\end{table}

\section{Discussion}
\label{discussion}
Most of the uncertainty in system mass is from the parallax error. Further
improvements in determining this parameter will improve the accuracy of system
mass considerably beyond the present $\pm 7.5\,\%$.\\
We also want to get a first
estimate of the components' masses in order to judge the probability for a
substellar object in the system. This cannot be done empirically from our
data, because there are no published radial velocities for
\object{Gliese~866}. Instead we use the mass-luminosity relation given by
Henry \& McCarthy (\cite{Henry93}) as

\begin{equation}
\log \frac{M}{M_{\odot}}\pm 0.067 = -0.1668M_{\mathrm{K}} + 0.5395
\label{mass-k}
\end{equation}
  
for absolute K magnitudes $9.81\le M_{\mathrm{K}} < 7.70$.\\
This approach is only valid if there is no large shift between the
spectral energy distributions of the components. To check this
we consider the observations taken at other wavelengths. At $917\,\mathrm{nm}$
the flux ratio is comparable to that in the K-band and also to the flux ratios
in other NIR filters given by L90 (Table\,3 therein). This supports the
conclusion of L90 that both (visual) components nearly have the same spectral
type and effective temperature. At $845\,\mathrm{nm}$ L90 have measured a
higher flux ratio of $I_{\mathrm{B}}/I_{\mathrm{Aa}} = 0.83\pm 0.1$, but this
 wavelength lies at
the edge of a strong TiO absorption feature (L90, Fig.\,8 therein), so an
extrapolation of flux ratios into the visible is not straightforward.
Furthermore Henry et al.\,(\cite{Henry99}) have
observed \object{Gliese~866} with the F583W filter of the HST. The resultant
flux ratio in V is $I_{\mathrm{B}}/I_{\mathrm{Aa}} = 0.69\pm 0.06$ and thus
again close to
the values in the near infrared. The fact that the flux ratio
$I_{\mathrm{B}}/I_{\mathrm{Aa}}$
is nearly constant over a large range of wavelengths indicates that
all {\it three} components of the \object{Gliese~866} system have similar
effective temperatures and thus similar masses. This idea is further 
supported by the combined spectrum of \object{Gliese~866} (L90, Fig.\,8
therein) that shows the deep molecular absorptions of an M5.5 star.\\
The apparent magnitude of
the system is $K = (5.56\pm 0.02)\,\mathrm{mag}$ (Leggett\,\cite{Leggett92} and
references therein). Combined with the distance given above the absolute
system magnitude is $(7.87\pm 0.04)\,\mathrm{mag}$. The K-band flux ratios
given in Table\,\ref{speckle-results} result in a mean value of
$I_{\mathrm{B}}/I_{\mathrm{Aa}} = 0.57\pm 0.01$. We take the components of
the spectroscopic pair
\object{Gliese~866}~Aa to be equally bright. Because we have only given
qualitative arguments for this assumption, we use an error reflecting this
uncertainty: $I_{\mathrm{a}}/I_{\mathrm{A}} = 1.0\pm 0.5$.
This yields the components' absolute K magnitudes:

\begin{eqnarray}
M_{\mathrm{K}}(A) & = & M_{\mathrm{K}}(a) = (9.11\pm 0.32)\,\mathrm{mag}
 \nonumber \\
M_{\mathrm{K}}(B) & = & (8.98\pm 0.05)\,\mathrm{mag}
\label{k-mag}
\end{eqnarray}

The resulting masses from Eq.\,\ref{mass-k} then are:

\begin{eqnarray}
M_{\mathrm{A}} & = & M_{\mathrm{a}} = (0.105\pm 0.021) M_{\odot} \nonumber \\
M_{\mathrm{B}} & = & (0.110\pm 0.018) M_{\odot}.
\label{masses}
\end{eqnarray}

The given uncertainties originate from the error of the K magnitudes
(Eq.\,\ref{k-mag}) and the error of the mass-luminosity relation
itself (Eq.\,\ref{mass-k}).
The sum of these masses is $M_{\mathrm{Sys}} = 0.320\pm 0.035\,M_{\odot}$ and
is thus within the uncertainties consistent with the dynamical system mass 
$M_{\mathrm{Sys, dyn}} = 0.336\pm 0.026\,M_{\odot}$ derived above.\\

\section{Summary}
\label{summary}
From a new determination of the visual orbit we have given an improved 
determination of the system mass in \object{Gliese~866}.
With simplifying assumptions we have given estimates for the components'
masses using the mass-luminosity relation by Henry \& McCarthy
(\cite{Henry93}). The sum of the components' masses estimated in this way is
consistent with the dynamically obtained system mass.
We conclude that there is no substellar object in the triple system 
\object{Gliese~866} despite the fact that the total system mass is only
$0.34\,M_{\odot}$.

\begin{acknowledgements}
We thank Rainer K\"ohler very much for providing his software package
''speckle'' for the reduction of 2D speckle-interferometric data.
Mark McCaughrean has contributed the procedure to calibrate pixel scale and
detector orientation with the Trapezium cluster which largely improved
the relative astrometry. William Hartkopf has provided orbits and
ephemerides for several visual binaries that we also used for determination
of pixel scale and detector orientation. We are grateful to Patrice Bouchet
for carrying out the observation at 16 August 1995 and to Jean-Luc Beuzit
for the observation at 25 August 1997.
\end{acknowledgements}

\end{document}